\documentclass[letter]{emulateapj}
\usepackage{apjfonts,psfig,epsfig}
\newcounter{fred}

\lefthead{Smith, G.P., et al.}
\righthead{Non--detection of the z=10 Galaxy}

\begin{document}

\def\ls{\mathrel{\hbox{\rlap{\hbox{\lower4pt\hbox{$\sim$}}}\hbox{$<$}}}}
\def\gs{\mathrel{\hbox{\rlap{\hbox{\lower4pt\hbox{$\sim$}}}\hbox{$>$}}}}

\title{Optical and Infrared Non--detection of the z=10 Galaxy 
Behind Abell~1835}

\author{
Graham P.\ Smith,$\!$\altaffilmark{1}
David J.\ Sand,$\!$\altaffilmark{1} 
Eiichi Egami,$\!$\altaffilmark{2}
Daniel Stern,$\!$\altaffilmark{3} and 
Peter R.\ Eisenhardt$\!$\altaffilmark{3}
}

\altaffiltext{1}{Department of Astronomy, California Institute of
                 Technology, Mail Code 105--24, Pasadena, CA 91125. --
                 Email: gps@astro.caltech.edu}

\altaffiltext{2}{Steward Observatory, University of Arizona, 933 North
                 Cherry Avenue, Tucson, AZ 85721.}

\altaffiltext{3}{Jet Propulsion Laboratory, California Institute of
                 Technology, 4800 Oak Grove Drive, MS 169--327,
                 Pasadena, CA 91109.}

\setcounter{footnote}{4}

\begin{abstract}

Gravitational lensing by massive galaxy clusters is a powerful tool
for the discovery and study of high redshift galaxies, including those
at $z{\ge}6$ likely responsible for cosmic re--ionization.  Pell\'o et
al.\ recently used this technique to discover a candidate
gravitationally magnified galaxy at $z{=}10$ behind the massive
cluster lens Abell~1835 ($z{=}0.25$).  We present new Keck (LRIS) and
\emph{Spitzer Space Telescope} (IRAC) observations of the $z{=}10$
candidate (hereafter \#1916) together with a re-analysis of archival
optical and near-infrared imaging from the \emph{Hubble Space
  Telescope} and VLT respectively.  Our analysis therefore extends
from the atmospheric cut-off at ${\lambda}_{\rm
  obs}{\simeq}0.35{\mu}{\rm m}$ out to ${\lambda}_{\rm
  obs}{\simeq}5{\mu}{\rm m}$ with \emph{Spitzer}/IRAC.  The $z{=}10$
galaxy is not detected in any of these data, including an independent
reduction of Pell\'o et al.'s discovery $H$- and $K$-band imaging.  We
conclude that there is no statistically reliable evidence for the
existence of \#1916.  We also assess the implications of our results
for ground-based near-infrared searches for gravitationally magnified
galaxies at $z{\gs}7$.  The broad conclusion is that such experiments
remain feasible, assuming that space-based optical and mid-infrared
imaging are available to break the degeneracy with low redshift
interlopers (e.g.\ $z{\sim}2{-}3$) when fitting spectral templates to
the photometric data.

\end{abstract}


\section{Introduction}\label{intro}

Observations of distant QSOs (Becker et al.\ 2001; Fan et al.\ 2002)
and the cosmic microwave background (Kogut et al.\ 2003) together
suggest that the universe was re-ionized somewhere between
$z{\simeq}6$ and $z{\simeq}20$.  Searching for the sources of
re-ionizing photons is currently an intense observational effort.
Most searches naturally concentrate on luminous systems, i.e.\ QSOs
and luminous galaxies, at $z{\sim}6{-}8$ as these should be easier to
detect than less luminous and more distant objects.  However QSOs
likely produced insufficient photons to accomplish re-ionization alone
(Fan et al.\ 2001; Barger et al.\ 2003), and the same may be true of
luminous ($L{\gs}0.3L^{\star}_{z{=}3.8}$) galaxies based on small
samples from the \emph{Hubble Space Telescope} Ultra Deep Field
(hereafter \emph{HST} UDF; Bouwens et al.\ 2004; Bunker et al.\ 2004;
Yan et al.\ 2004).  This raises the important possibilities that
re-ionization either occurred much earlier, or that the bulk of the
re-ionizing photons were emitted by sub-luminous galaxies,
i.e.\ $L{\ls}0.1L^{\star}$.

The UDF studies operate close to the detection threshold of the
deepest optical/near-infrared imaging available.  It is therefore
difficult to envisage substantial progress in the detection of more
remote and/or less luminous galaxies via deep imaging of ``blank
fields'' with the current generation of telescopes.  With the advent
of the \emph{James Webb Space Telescope} still some years ahead, the
magnifying power of massive galaxy cluster lenses is therefore a much
needed boost for the discovery power of \emph{HST} and large
ground-based telescopes.  Indeed, the galaxy redshift record has been
broken on several occasions with the help of the gravitational
magnification of distant galaxies by foreground galaxy clusters
(Mellier et al.\ 1991; Franx et al.\ 1997; Hu et al.\ 2002; Kneib et
al.\ 2004).  The faint end of the luminosity function of
Lyman-${\alpha}$ emitters at $z{=}5$ has also been constrained with
the help of gravitational lensing (Santos et al.\ 2004; Ellis et al.\
2001).  Extension of these techniques to $z{\gs}7$ is therefore an
important element of observational studies of cosmic re-ionization .

Pell\'o et al.\ (2004 -- hereafter P04) reported a gravitationally
magnified (${\mu}{\sim}25{-}100$) galaxy at $z{=}10$ (hereafter
\#1916, following P04's nomenclature) behind the foreground galaxy
cluster A\,1835 ($z{=}0.25$).  This interpretation is based on
non-detection in optical imaging from the ground ($3{\sigma}$ limits
in a $0.6''$ diameter aperture: $V{\ge}27.4$, $R{\ge}27.5$,
$I{\ge}26.9$) and space ($3{\sigma}$ limit in a $0.2''$ diameter
aperture: $R_{702}{\ge}27.2$), and the shape of the continuum at
${\lambda}_{\rm obs}{\ge}1{\mu}{\rm m}$ (using a $1.5''$ aperture:
$(J{-}H){\ge}0.6$, $(H{-}K){=}{-}0.5{\pm}0.4$) which is reminiscent of
the Lyman-break selection technique (Steidel et al.\ 1996).  P04
corroborated the putative Lyman-break redshifted to ${\lambda}_{\rm
obs}{\simeq}1.3{\mu}{\rm m}$ with an emission line at ${\lambda}_{\rm
obs}{=}1.3375{\mu}{\rm m}$ with integrated flux of
$(4.1{\pm}0.5){\times}10^{-18}{\rm erg\,cm^{-2}\,s^{-1}}$, which they
interpret as Lyman-${\alpha}$.  Lower redshift interpretations of the
line ([OII] at $z{=}2.59$; [OIII] at $z{=}1.68$; H${\alpha}$ at
$z{=}1.04$) were discarded by P04 largely on the basis of the low
probability of solutions at $z{\ls}7$ when fitting synthetic spectral
energy distributions (SEDs) to their photometric data.  The most
likely of these lower-redshift solutions ($z{=}2.59$) was further
excluded on the basis of the dust extinction required to fit the
photometric data ($A_V{\ge}2$), and the absence of doublet structure
in the observed emission line.

Based solely on the photometry (optical non-detection, red $(J{-}H)$
and blue $(H{-}K)$ colors) the $z{=}10$ interpretation of \#1916 is
plausible.  However P04's preference for this solution over the lower
redshift alternatives was controversial from the outset.  For example,
the emission line does not have the characteristic P-Cygni profile of
Lyman-${\alpha}$, and the different photometric apertures adopted in
the optical ($0.6''$ -- smaller than the ground-based seeing disc) and
near-infrared ($1.5''$ -- $3{\times}$ the seeing disk) may suppress
the likelihood of lower-redshift solutions when fitting synthetic
SEDs.  Bremer et al.\ have also suggested that \#1916 may either not
exist, or be intrinsically variable, based on their non-detection in
the $H$-band with NIRI on Gemini-North, $H(3{\sigma}){>}26$, in
contrast to P04's $4{\sigma}$ detection of $H{=}25.00{\pm}0.25$ with
ISAAC on VLT.  The spectroscopic identification of \#1916 is also in
doubt.  Weatherley et al.\ (2004) re-analyzed P04's spectroscopic
data and failed to detect the emission line at ${\lambda}_{\rm
obs}{=}1.3375{\mu}{\rm m}$, citing spurious positive flux arising from
variable hot pixels in the ISAAC array as the likely source of the
discrepancy.

A\,1835 has been used previously as a gravitational telescope, for
example targeting sub-millimeter galaxies (hereafter SMGs; e.g.\
Smail et al.\ 2002) and galaxies with extremely red
optical/near-infrared colors (Smith et al.\ 2002).  One of the
galaxies detected in these surveys, SMMJ\,14011${+}$0252, lies at
$z{=}2.56$ and suffers an estimated extinction of
$1.8{\ls}A_V{\ls}6.5$ (Ivison et al.\ 2000).  Bearing in mind recent
discovery of galaxy groups at $z{\sim}2{-}3$ associated with
gravitationally lensed SMGs (Kneib et al.\ 2004; Borys et al.\ 2004),
and the strong clustering of SMGs (Blain et al.\ 2004), \#1916 is
plausibly at a similar redshift to SMMJ\,14011${+}$0252, and may also
be obscured by dust.  Further circumstantial evidence for a lower
redshift interpretation of \#1916 comes from Richard et al.\ (2003)
who used the same spectroscopic data as presented by P04 to discover a
strongly reddened star-forming galaxy at $z{=}1.68$.  This redshift
coincides with the [OIII] interpretation of PO4's putative emission
line at ${\lambda}_{\rm obs}{=}1.3375{\mu}{\rm m}$.

In this paper, we address three questions: (i) is \#1916 at
$z{\sim}2{-}3$?; (ii) is \#1916 intrinsically variable?; (iii) does
\#1916 exist?  These tests exploit new optical and mid-infrared
observations using the Keck-I 10-m telescope and the \emph{Spitzer
  Space Telescope} respectively, plus an independent reduction of
P04's $H$- and $K$-band imaging data from VLT.  Throughout this
article we assume that the emission line at ${\lambda}_{\rm
  obs}{=}1.3375{\mu}{\rm m}$ is a false detection (Weatherley et
al.\ 2004).  In \S\ref{data} we present the new Keck and
\emph{Spitzer} data, explain in detail the re-reduction of the
archival VLT/ISAAC data, and summarize the archival \emph{Hubble Space
  Telescope (HST)} data.  Then in \S\ref{analysis} we describe the
analysis and key results, focusing on the three questions posed above;
this section closes with a summary of the current observational status
of \#1916.  Finally, we discuss the implications of our results for
future ground-based near-infrared searches for galaxies at $z{\gs}7$
(\S\ref{discuss}) and summarize our conclusions in
(\S\ref{conclusions}).

We assume $H_0{=}65{\rm km\,s^{-1}Mpc^{-1}}$, ${\Omega}_{\rm M}{=}0.3$
and ${\Omega}_{\rm \Lambda}{=}0.7$ throughout.  Unless otherwise
stated all error bars are at $1{\sigma}$ significance; photometric
detection limits are at $3{\sigma}$ significance; upper and lower
limits on colors are based on $3{\sigma}$ detection thresholds in the
non-detection filter.  Magnitudes are stated in the AB system;
conversion between the AB and Vega systems for the specific filters
used in this paper are as follows:- ${\Delta}_B{=}B_{\rm AB}{-}B_{\rm
Vega}{=}{-}0.1$, ${\Delta}_V{=}0.1$, ${\Delta}_R{=}0.2$,
${\Delta}_{F702W}{=}0.3$, ${\Delta}_I{=}0.5$, ${\Delta}_J{=}0.9$,
${\Delta}_H{=}1.4$, ${\Delta}_K{=}1.9$, ${\Delta}_{3.6{\mu}{\rm
m}}{=}2.8$, ${\Delta}_{4.5{\mu}{\rm m}}{=}3.2$.

\section{Observations and Data Reduction}\label{data}

%
%
\setcounter{table}{0}
\begin{table}
\begin{center}
\caption{
Summary of Photometry
\label{table:imaging}
}
\begin{tabular}{llcccc}
\hline
Filter & Telescope/          & FWHM   & \multispan2{\hfil Aperture Photometry$^a$\hfil}\cr
       & Instrument          & ($''$) & Pell\'o et al.$^b$              & This Paper \cr
\hline
$V$    & CFHT/12k            & 0.76   & ${\ge}27.5$\,($0.6''$)           & {...} \cr
$R$    & CFHT/12k            & 0.69   & ${\ge}27.6$\,($0.6''$)           & {...} \cr
$R_{702}$ & \emph{HST}/WFPC2 & 0.17   & ${\ge}27.2\,(0.2'')^c$             & ${\ge}27.0\,(0.5'')$ \cr
$I$    & CFHT/12k            & 0.78   & ${\ge}26.0$\,($0.6''$)           & {...} \cr
$J$    & VLT/ISAAC           & 0.65   & ${\ge}25.6$\,($1.5''$)           & {...} \cr
$H$    & VLT/ISAAC           & 0.50   & $25.00{\pm}0.25$\,($1.5''$)      & ${\ge}25.0\,(1.5'')$ \cr
$K$    & VLT/ISAAC           & 0.38   & $25.51{\pm}0.36$\,($1.5''$)      & ${\ge}25.0\,(1.5'')$ \cr
$3.6{\mu}{\rm m}$ & \emph{Spitzer}/IRAC & 1.7 & {...}                    & ${\ge}24.3\,(5.1'')$ \cr
$4.5{\mu}{\rm m}$ & \emph{Spitzer}/IRAC & 1.7 & {...}                    & ${\ge}24.3\,(5.1'')$ \cr
\hline
\end{tabular}
\tablenotetext{a}{ 
  Each number in parentheses is the diameter of the aperture used for
  the respective photometric measurements.  In \S\ref{discuss}
  P04's optical non-detections are re-scaled to a photometric
  aperture of $2''$ diameter (${\sim}3$ times the seeing disc):
  $V{\ge}26.2$, $R{\ge}26.3$, $I{\ge}24.7$.
}
\tablenotetext{b}{
  We convert all of P04's optical detection limits and their
  $1{\sigma}$ $J$-band detection to $3{\sigma}$ limits.
}
\tablenotetext{c}{
  P04 do not state whether their $R_{702}$ detection limit is in the
  Vega or AB system.  We have assumed the former and converted it to
  AB in this table.  P04 also do not explain how they reduced the
  WFPC2 data, specifically whether the final pixel scale was different
  from the native $0.0996''/{\rm pix}$ of the WFC detectors.  In this
  table we have assumed that the four pixels over which this detection
  limit is measured (see P04 \S2.1) subtend a solid angle of
  $0.0996''{\times}0.0996''$.  Given these
  uncertainties and the absence of this detection limit from P04's
  Table~1 and Fig.~3, we ignore P04's limit when attempting to
  reproduce their photometric redshift results in
  \S\ref{discuss}.
}
\end{center}
\end{table}

We describe new and archival observations of \#1916 in order of
increasing wavelength, spanning the observed optical, near-infrared
and mid-infrared: spectroscopy with LRIS on Keck-I
($0.35{\le}{\lambda}_{\rm obs}{\le}0.95{\mu}{\rm m}$); imaging with
WFPC2 on-board \emph{HST} (${\lambda}_{\rm obs}{=}0.7{\mu}{\rm m}$);
near-infrared imaging with ISAAC on ESO's VLT (${\lambda}_{\rm
  obs}{=}1.6{\mu}{\rm m}$ and $2.2{\mu}{\rm m}$); IRAC/\emph{Spitzer}
observations at ${\lambda}_{\rm obs}{=}3.6{\mu}{\rm m}$ and
$4.5{\mu}{\rm m}$.

The detection of any optical flux from \#1916 would eliminate the
$z{=}10$ interpretation (e.g.\ Stern et al.\ 2000).  In contrast,
optical non-detection would have several alternative interpretations,
including: a galaxy at $z{=}10$ as per P04; a dusty galaxy at
$z{\sim}2{-}3$; non-existence of \#1916.  The mid-infrared
observations (\S\ref{spitzer}) should therefore help to constrain the
amount of energy re-radiated by dust in the $z{\sim}2{-}3$
interpretation, and our re-reduction of P04's VLT data also helps to
clarify the possibility that \#1916 may not exist, or may be variable
(Bremer et al.\ 2004).  The Keck and \emph{Spitzer} data described in
this section were collected and analyzed in parallel with those
presented by Bremer et al.\ (2004).

\begin{figure*}
\centerline{
\psfig{file=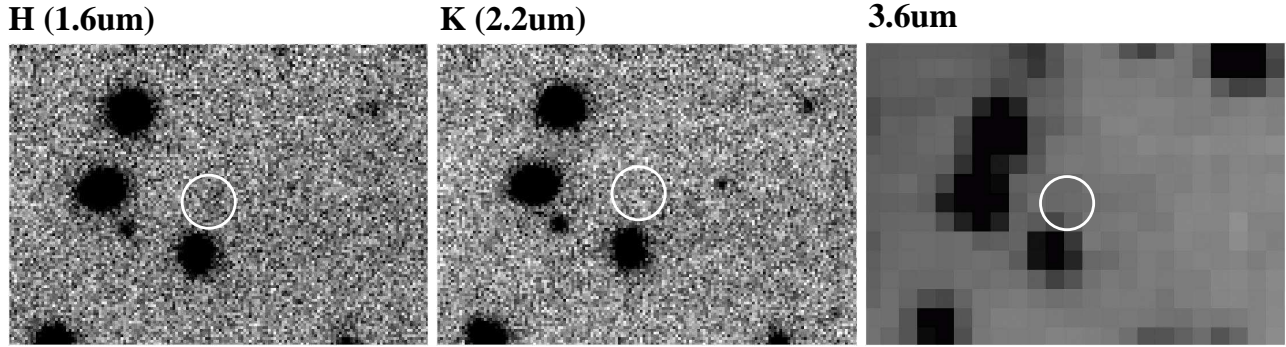,width=170mm,angle=0}
}
\caption{Infrared images of the $z{=}10$ candidate at $1.6, 2.2, {\rm
    and}\, 3.6{\mu}{\rm m}$ respectively.  The two left panels are based
  on our independent re-reduction of P04's ISAAC data described in
  \S\ref{isaac}.  The white circles mark the position of \#1916 from
  P04 -- there is no obvious sign of flux in any of these panels.
  Formal $3{\sigma}$ detection limits are: $H{\ge}25$, $K{\ge}25$,
  $F{\le}0.75{\mu}{\rm Jy}$.  North is up and East left.  Each panel is
  $23''{\times}16''$.
\label{fig:nir}
}
\end{figure*}

\subsection{Keck Spectroscopy}\label{keck}

As part of a broad effort to secure spectroscopic redshifts of
gravitational arcs spanning several observational programs (Smith et
al.\ 2001, 2002, 2005; Sand et al.\ 2002, 2004, 2005; Edge et
al.\ 2003; Sharon et al.\ 2005), we observed A\,1835 with the Low
Resolution Imager Spectrograph (LRIS; Oke et al.\ 1995) in multi-slit
mode on the Keck-I 10-m telescope\footnote{The W.\ M.\ Keck
  Observatory is operated as a scientific partnership among the
  California Institute of Technology, the University of California,
  and NASA.} on UT 2004 March 29.  A single mask was observed,
including a slit targeting \#1916.  The purpose of this slit was to
search for line emission in the $0.35{\le}{\lambda}_{\rm
  obs}{\le}0.95{\mu}{\rm m}$ wavelength range.  For example, if \#1916
does indeed lie at $z{\simeq}2.6$ (\S\ref{intro}), then
Lyman-${\alpha}$ may be detectable at ${\lambda}_{\rm
  obs}{\simeq}0.44{\mu}{\rm m}$.

The observations totaled 3.6-ks, split into two exposures, using the
D560 dichroic with the 400/8500 grating and the 400/3400 grism. On the
red side the spectral dispersion was 1.86\AA/pixel with a pixel scale
of $0.214''$/pixel and on the blue side the spectral dispersion was
1.09\AA/pixel with a pixel scale of $0.135''$/pixel.  Overhead
conditions were moderate (FWHM${\simeq}1''$), and probably not
photometric, however a flux calibration was obtained using the
spectrophotometric standard star HZ44 (Oke et al.\ 1990).  The data
were de-biased, flat-fielded, sky-subtracted, extracted and
calibrated in a standard manner within {\sc iraf}\footnote{{\sc iraf}
  is distributed by the National Optical Observatories, which is
  operated by the Association of Universities for Research in
  Astronomy, Inc.\ (AURA) under cooperative agreement with the
  National Science Foundation.}.

No flux at observed optical wavelengths has yet been detected at the
position of \#1916 (P04; Lehnert et al.\ 2004; \S\ref{hubble}).  We
therefore did not expect to detect any continuum emission, and
concentrated instead on searching for faint emission lines of large
equivalent width.  Visual inspection of the reduced 2D data revealed
neither continuum nor line emission.  To estimate the sensitivity
limit we extracted a 1D trace from the center of the slit
corresponding to the full width of the seeing disk, and estimated the
$3{\sigma}$ detection limit per 5-A spectral resolution element to be
${\sim}4.5{\times}10^{-19}{\rm erg\,s^{-1}\,cm^{-2}}$ at
${\lambda}_{\rm obs}{=}0.44{\mu}{\rm m}$ -- i.e.\ the observed
wavelength at which Lyman-$\alpha$ would be found if \#1916 is at
$z{=}2.6$.

\subsection{Archival \emph{Hubble Space Telescope} Imaging}\label{hubble}

A\,1835 has also been observed through the F702W filter with the WFPC2
camera on-board \emph{HST}\footnote{Based in part on observations with
  the NASA/ESA \emph{Hubble Space Telescope} obtained at the Space
  Telescope Science Institute, which is operated by the Association of
  Universities for Research in Astronomy, Imc., under NASA contract
  NAS 5-26555.}.  We refer the reader to Smith et al.\ (2005) for
details of these data and their reduction.  P04 do not detect \#1916
in these data (Table~\ref{table:imaging}), although they neither
explain how they reduced the data nor how the detection limit was
calculated.  Here, we use Smith et al.'s (2005) reduced frame which
has a pixel-scale of $0.0498''/{\rm pixel}$ after drizzling.  To
simplify the analysis, we re-bin the data back to the original
pixel-scale of $0.0996''/{\rm pixel}$ to minimize the impact of
pixel-to-pixel correlations in the background noise when estimating
the sensitivity limit of the data.

Visual inspection reveals no obvious flux at the position of \#1916 in
these data.  To quantify this non-detection, we follow the same
procedure as P04 -- we measured the background noise in apertures
placed randomly into blank sky regions of both frames near to the
position of \#1916.  We ensure consistency with other wavelengths by
matching the diameter of these apertures to a diameter $3{\times}$
that of the seeing disk, i.e.\ $0.5''$.  We obtain a $3{\sigma}$
sensitivity limit in that aperture of $R_{702}{=}27.0$
(Table~\ref{table:imaging}).

\subsection{Archival VLT/ISAAC Imaging}\label{isaac}

Deep near-infrared imaging of A\,1835 was obtained by P04 with the
ISAAC $1024{\times}1024$ Hawaii Rockwell array on ESO's 8-m
VLT\footnote{Based in part on observations collected with the ESO
  VLT-UT1 Antu Telescope.} in February 2003.  We reduced independently
the $H$- and $K$-band data using standard {\sc iraf} tasks, paying
careful attention to rejection of cosmetic defects in the ISAAC array,
including bad pixels, and to conserving the noise properties of the
data.  We detail key features of the data reduction below.

\setcounter{fred}{0}
\begin{list}{(\roman{fred})}{\usecounter{fred}\setlength{\itemindent}{0mm}\setlength{\labelwidth}{15mm}\setlength{\labelsep}{2mm}\setlength{\leftmargin}{7mm}} 
\setlength{\itemsep}{1.0mm}

\item Flat-fielding and sky-subtraction were combined into a single
  step using a median of the eight frames temporally adjacent to each
  science frame.  We refer to this step as ``flat-fielding'', and to
  the rolling temporal median frames as ``sky-flats''.

\item Flat-fielding was performed twice.  First on the dark-subtracted
  frames which were then registered and averaged to produce a
  first-pass reduced frame.  This frame was then used to mask out flux
  from identified sources from the individual dark-subtracted frames,
  and these masked frames were then used to construct the sky-flats in
  a second-pass reduction.  This approach minimizes the loss of flux
  from objects with angular extents comparable with the size of the
  dither pattern.  This is important when searching for faint objects
  along lines-of-sight through the crowded cores of rich galaxy
  clusters because the light from bright cluster galaxies 
  effectively form a spatially varying background against which the
  faint sources are detected.  The goal of the second-pass flat-field
  is to conserve this ``background''.

\item Independent bad pixel masks were made by sigma-clipping both
  the darks and the sky-flats.  The former identifies $22,020$ pixels
  ($2.1\%$ of the total array) as static, i.e.\ ``bad'', and the
  latter identifies the same $22,020$ pixels plus an additional
  $12,373$ pixels (a further $1.2\%$ of the total array) as bad.  The
  latter mask was adopted as the fiducial bad pixel mask.

\item Detector bias residuals were removed by subtracting the median
  along rows in individual flat-fielded frames after masking
  identified sources, in a manner similar to that described by Labb\'e
  et al.\ (2003).

\item The individual frames were integer pixel aligned.  This has the
  important benefit of minimizing pixel-to-pixel correlations in the
  noise properties of each frame and thus the final stacked frame.
  Calculation of the background noise is therefore simplified relative
  to a reduction scheme based on sub-integer pixel alignment of the
  individual frames.

\item No frames were rejected when making the final combination of the
  reduced, aligned frames.  Two versions of the final stacked frame
  were made: a straight average and a weighted average -- the weight
  of an individual frame was proportional to $({\sigma}{\times}{\rm
    rms})^{-2}$, where ${\sigma}$ is the FWHM of the seeing disk, and
  rms is the root mean square per pixel of the noise in each frame.
  The weighted version of the final frame has slightly better image
  quality than the straight average.  We therefore adopt it for the
  analysis described below.

\end{list}

The final reduced $H$- and $K$-band frames have seeing of ${\rm
  FWHM}{=}(0.45{\pm}0.01)''$ and ${\rm FWHM}{=}(0.34{\pm}0.02)''$
respectively.  Photometric calibration was achieved with the standard
star observations that were interspersed with the science observations
as part of P04's original program.  We show extracts from the reduced
$H$- or $K$-band frames in Fig.~\ref{fig:nir}.  Visual inspection of
both the fits frames and Fig.~\ref{fig:nir} reveals no obvious flux at
the location of \#1916 in either the $H$- or $K$-band.  Following P04
we again randomly insert $1.5''$ diameter apertures (roughly 3 times
the seeing disc) into blank sky regions near to the position of
\#1916, obtaining $3{\sigma}$ sensitivity limits in this aperture of:
$H{=}25.0$ and $K{=}25.0$.

\subsection{\emph{Spitzer}/IRAC Imaging}\label{spitzer}

A\,1835 was observed with the InfraRed Array Camera (IRAC; Fazio et
al.\ 2004) on-board \emph{Spitzer}\footnote{This work is based in part
  on observations made with the Spitzer Space Telescope, which is
  operated by the Jet Propulsion Laboratory, California Institute of
  Technology under a contract with NASA.} on UT 2004 January 16 in the
3.6, 4.5, 5.8 and 8.0${\mu}{\rm m}$ channels.  Here we discuss the two
shortest wavelength, more sensitive, observations.  Twelve and
eighteen 200-second exposures were accumulated at $3.6{\mu}{\rm m}$
and $4.5{\mu}{\rm m}$ respectively, using the small-step cycling
dither pattern.  The Basic Calibrated Data (BCD) were combined using
custom routines to produce the final stacked frame with a pixel scale
of $0.6''/{\rm pixel}$.  We re-binned the data back to the original
pixel scale of $1.2''/{\rm pixel}$ to eliminate correlations in the
background noise.

Visual inspection of the final frames again indicates that there is no
flux at the position of \#1916 (Fig.~\ref{fig:nir}).  To quantify this
non-detection we follow the same procedure as P04, as described in
\S\ref{hubble}.  We used $5.1''$ diameter apertures, i.e.\ $3{\times}$
the seeing disk of the IRAC observations to obtain $3{\sigma}$
sensitivity limits of: $F(3.6{\mu}{\rm m}){=}0.75{\mu}{\rm Jy}$ and
$F(4.5{\mu}{\rm m}){=}0.75{\mu}{\rm Jy}$ respectively.

\section{Analysis and Results}\label{analysis}

The objective of this section is to answer the three questions posed
in \S\ref{intro}: (i) is \#1916 at $z{\sim}2{-}3$?; (ii) is \#1916
intrinsically variable?; (iii) does \#1916 exist?  Preliminary
inspection of the data in \S\ref{data} indicates that no flux is
detected at the position of \#1916 at any wavelength to date.
Combining this with Bremer et al.'s more sensitive non-detection of
$H(3{\sigma}){>}26.0$, it is tempting to leap to the third question
and reply ``no''.  We adopt a more conservative approach.

\subsection{Is \#1916 at $z{\sim}2{-}3$?}\label{q1}

This test concentrates on the optical data because the detection of
any flux shortward of the putative Lyman limit of a galaxy at
$z{\simeq}10$ would immediately discount that interpretation.  The red
observed optical/near-infrared spectral energy distribution described
by P04 could then be naturally explained by a dusty galaxy at
$z{\sim}2{-}3$, perhaps associated with the SMGs that lie within
${\sim}30''$ (${\sim}200{-}300{\rm kpc}$ in projection at
$z{\sim}2{-}3$) of \#1916 (Ivison et al.\ 2000; Smail et al.\ 2005).

Our new non-detection of \#1916 with LRIS (\S\ref{keck}), coupled with
confirmation of P04's non-detection with \emph{HST}/WFPC2 and Lehnert
et al.'s non-detection in the $V$-band with VLT/FORS are mutually
consistent in the sense that no optical flux has been detected at this
position to date.  However these non-detections are consistent with
all of the following: $z{=}10$, extreme dust obscuration at
$z{\sim}2{-}3$, an intrinsically variable source, and non-existence.
The result of this test is therefore inconclusive.

\subsection{Is \#1916 Intrinsically Variable?}\label{q2}

The objective of this section is to test Bremer et al.'s (2004)
proposal that \#1916 is intrinsically variable.  If P04's photometry
($H{=}25.00{\pm}0.25$ and $K{=}25.51{\pm}0.51$) is reproducible using
our independent reduction of their near-infrared data, then the variable
hypothesis would be supported.  If not, then the idea that \#1916 does
not exist would gain credibility (\S\ref{q3}).

We attempt to reproduce P04's analysis using SExtractor (Bertin \&
Arnouts 1996).  SExtractor was configured to locate all sources with
at least 7 pixels that are ${\ge}0.75{\sigma}$ per pixel above the
background -- i.e.\ a signal-to-noise ratio of ${\gs}2$ per resolution
element, based on the $H$-band seeing disk of ${\rm
  FWHM}{=}0.45{\pm}0.01''$ (\S\ref{isaac}) and the $0.15''/{\rm pix}$
scale of the ISAAC pixels.  We also smoothed the data with a gaussian
filter that matched the FWHM of the observed point sources, i.e.\ a
gaussian of FWHM${=}3$ pixels.  In this configuration SExtractor
failed to detect a source at the position of \#1916.  We therefore
experimented with different smoothing schemes, both increasing and
decreasing the full width of the gaussian filter.  A ``detection'' was
only possible with the smallest available filter -- FWHM${=}1.5$
pixels, i.e.\ half the width of the seeing disk -- yielding
$H{=}25.3{\pm}0.6$.  Experimentation with block filters produced
similar results in that a ``detection'' was not possible with any of
the standard SExtractor block filters: $3{\times}3$, $5{\times}5$,
$7{\times}7$ pixels.  We also analyzed the $K$-band data in exactly
the same manner and failed to detect anything at the position of
\#1916 with any gaussian or block filter.

The $H$-band segmentation map produced when smoothing with the
FWHM${=}1.5$ pixel gaussian reveals that the ``detection'' is very
elongated, with a width of $1{-}2$ pixels and a length of ${\sim}5$
pixels.  The orientation of these pixels is consistent with the
orientation of \#1916 reported by P04.  It is important to stress that
the motivation for filtering data with a kernel that matches the
resolution element of the data is to suppress false detections.  The
collection of pixels identified by SExtractor at the position of
\#1916 was only ``detectable'' with a smoothing kernel that has a
linear scale half that of the resolution element of the data.  It is
therefore instructive to consider how many such ${\sim}2{\sigma}$
blobs exist within the ISAAC data.  In a single $1.5''$ diameter
aperture (i.e.\ matching that used for the photometry described above)
placed randomly in these $H$-band data, there is a $5\%$ chance of
detecting a $2{\sigma}$ noise fluctuation -- i.e.\ a spurious
detection.  However the ISAAC array ($1024{\times}1024{\rm pixels}$,
each pixel $0.15''{\times}0.15''$) contains of order $10^4$
independent photometric apertures of $1.5''$ diameter.  The $H$-band
frame therefore contains ${\sim}500$ noise fluctuations of $2{\sigma}$
significance.

Sadly, the only reasonable conclusion to draw from this analysis is
that \#1916 is not detected in our independent reduction of
P04's data.  We therefore place $3{\sigma}$ limits on the flux at this
position of: $H{\ge}25.0$ and $K{\ge}25.0$ (\S\ref{isaac}).  The only
wavelength at which two directly comparable observations are available
is in the $H$-band.  Combining our non-detection with that of Bremer
et al.\ (2004), we conclude that there is no evidence for variability
of \#1916, and that (if it exists) its $H$-band flux is fainter than
$H{=}26$ at $3{\sigma}$ significance (Bremer et al.\ 2004).

\begin{figure*}
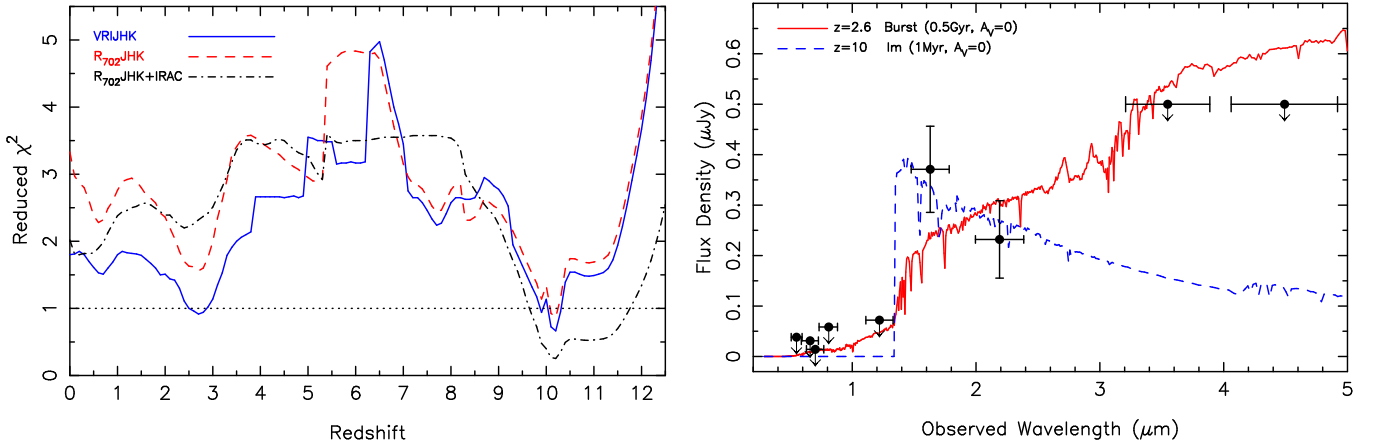

\centerline{
\psfig{file=f2a.ps,width=88mm,angle=0}
\hspace{2mm}
\psfig{file=f2b.ps,width=88mm,angle=0}
}
\caption{ 
  {\sc left}: Reduced ${\chi}^2$ as a function of photometric redshift
  and the filters used in the model fits.  The red dashed curve has
  been offset ${+}0.2$ in the $Y$-direction for clarity.  This panel
  demonstrates that P04's discovery photometry ($VRIJHK$) is
  degenerate between $z{\simeq}2.6$ and $z{=}10$ (blue solid curve).
  Adding in sensitive space-based detection thresholds from
  \emph{HST} and \emph{Spitzer}/IRAC (red dashed and black dot-dashed
  curves respectively) lifts this degeneracy.  {\sc right}: The
  best-fit spectral templates to the $R_{702}JHK$-band photometry at
  $z{=}2.6$ (red solid) and $z{=}10$ (blue dashed).  The IRAC
  detection thresholds are also marked to illustrate the power of
  these data to discriminate between low- and high-redshift
  interpretations of the shorter wavelength data.
\label{fig:photoz}
}
\end{figure*}

\subsection{Does \#1916 exist?}\label{q3} 

The results of the preceding two sections were derived from
non-detection of \#1916 across the broadest wavelength range to date:
$0.35{\le}{\lambda}_{\rm obs}{\le}5{\mu}{\rm m}$.  We now combine all
of these non-detections to address the question of whether \#1916
exists.  The data force us to conclude that there is no statistically
sound evidence that \#1916 exists.  The balance of probability is that
\#1916 was a false detection in P04's discovery observations.  New
observational data yielding statistically sound detections are
required before P04's claim that \#1916 is the most distant galaxy yet
discovered may be resurrected.  We consider this unlikely, but we hope
to be surprised by Pell\'o et al.'s forthcoming \emph{HST}
observations.

\section{Discussion}\label{discuss}

\subsection{The Near-infrared Non-detection}

We first discuss possible reasons for the difference between our
non-detection of \#1916 and P04's ${\sim}3{-}4{\sigma}$ near-infrared
detections.  We single out two data reduction steps (\S\ref{isaac})
for discussion -- the efficiency of bad pixel rejection and
the conservation of noise properties.

\subsubsection{Efficiency of bad pixel rejection} 

The efficiency with which bad pixels are identified can affect
estimates of the signal coming from faint sources -- if some bad
pixels are not identified, they could enhance the flux detected by
SExtractor.  In \S\ref{isaac} we used two different methods to
identify bad pixels, finding a small but potentially important
difference between the two methods.  To assess the impact of reduced
efficiency of bad pixel identification we made a mask image that
contained the 12,373 bad pixels identified only in the bad pixel
mask generated from the sky-flats.  We then made one copy of the mask
per science frame and integer-pixel shifted them to match the observed
dither pattern.  Finally we took the weighted average of these aligned
mask frames and summed the pixel counts in a $1.5''$ diameter aperture
centered on \#1916.  From this we concluded that 7 bad pixels
that are only identified in our ``sky-flat'' bad pixel maps fall
within the final photometric aperture.  We estimate that if not
identified and excluded from the analysis, these pixels could increase
the flux estimates by several tenths of a magnitude.

\subsubsection{Conservation of noise properties}

The approach to re-sampling (or not) the individual frames during the
data reduction process, especially the alignment of individual frames
affects the noise properties of the final stacked frame.  If the
individual frames are re-sampled, for example by sub-integer pixel
aligning them immediately prior to producing the final stacked frame,
then the noise in the stacked frame is correlated.  Such
pixel-to-pixel correlations are generally absent from integer-pixel
aligned data, thus simplifying the noise properties of the final
frame.  Neither sub-integer nor integer pixel alignment is
intrinsically correct.  The relevant issue is correct measurement of
the noise in each case -- this is critical to assess accurately the
statistical significance of sources detected close to the sensitivity
limit of the data.  Specifically, if the pixel-to-pixel correlations
in sub-integer pixel aligned data are not included in the error
analysis, then the noise is under-estimated and the statistical
significance of the detection over-estimated (Casertano et al.\ 2000).
We integer pixel aligned the individual frames in \S\ref{isaac} in
order to simplify the error analysis.  We now estimate by how much we
would have under-estimated the noise if we had sub-integer pixel
aligned the individual frames and then ignored the pixel-to-pixel
correlations when calculating noise level.  This is achieved by simply
sub-integer aligning the individual frames and re-combining them using
a weighted average.  Ignoring any resulting pixel-to-pixel
correlations, we obtain a $3{\sigma}$ threshold of $H{=}25.2$,
which is slightly fainter than our threshold of $H{=}25.0$.  

\smallskip

In summary, it is plausible that the difference between our
near-infrared non-detection and P04's detection of \#1916 using the
same raw data can at least in part be explained by the efficiency of
bad pixel identification and treatment of correlated noise.

\subsection{Implications for Future Work}

We now discuss the implications of our results for future work,
focusing on the feasibility of searches for gravitationally magnified
stellar systems at $z{\gs}7$ using ground-based near-infrared data.
We begin by noting that, had their photometry been reliable, P04's
original $z{=}10$ interpretation of \#1916 was plausible, based solely
on the photometric data.  We therefore adopt P04's optical and
near-infrared photometry as being representative of what may be
expected from similar future experiments -- i.e.\ optical
non-detections in several filters based on a few hours of observations
with a 4-m class telescope, a non-detection in a single red optical
filter with \emph{HST}, ${\sim}3{-}4{\sigma}$ detections in two
near-infrared filters, and possibly observations with
\emph{Spitzer}/IRAC.  Specifically, we investigate the degeneracy
between $z{\gs}7$ interpretations of such datasets with lower redshift
alternatives, and how such degeneracies might be broken.

We use Version~1.1 of {\sc hyper-z}\footnote{ Available from
http://webast.ast.obs-mip.fr/hyperz} (Bolzonella et al.\ 2000) to fit
standard Bruzual \& Charlot (1993) single stellar population models
(Burst, E, Sa, Sc, Im) to the photometric data.  We assume a Calzetti
et al.\ (2000) extinction law, allow dust extinction in \#1916 to lie
in the range $0{\le}A_V{\le}4$, adopt $E(B-V)=0.03$ for extinction
within the Milky Way (Schlegel et al.\ 1998), and use Madau's (1995)
prescription for absorption by the inter-galactic medium.  In each
case described below, if we fit the models to all available
photometric information across the full redshift range
($0{\le}z{\le}11$), then we obtain ${\chi}^2{\ls}1$ (all ${\chi}^2$
values are quoted per degree of freedom) for all redshifts beyond
$z{\simeq}7$.  This is because the models are defined to have zero
flux short-ward of the Lyman limit (${\lambda}_{\rm
rest}{=}0.0912{\mu}{\rm m}$).  At $z{\gs}6$, the models also have
negligible flux short-ward of Lyman-${\alpha}$ (${\lambda}_{\rm
rest}{=}0.1215{\mu}{\rm m}$) due to the Lyman forests and Gunn
Peterson absorption.  These two spectral features are progressively
redshifted long-ward of the observed optical filters at high
redshift, rendering the shorter wavelength detection limits irrelevant
to the fits.  The goodness of fit is therefore systematically
over-estimated at high redshift if all photometric information is
included in the fit across the full redshift range.  We therefore fit
the models to the data in a series of redshift chunks -- only the
photometric information that lies longward of red-shifted
Lyman-${\alpha}$ is considered in each redshift chunk.  The results
described below are insensitive to whether the Lyman limit is
substituted for Lyman-${\alpha}$.

First, we fit the models to P04's $VRIJHK$-band ground-based
photometry (i.e.\ the $VRIJ$-band detection limits and the $HK$-band
detections listed in column 4 of Table~\ref{table:imaging}).  Note
that P04 used different sized apertures for the near-infrared
($1.5''$) and optical ($0.6''$) photometry respectively.  We find two
acceptable solutions: $z{\simeq}2.6$ and $z{\simeq}10$
(Fig.~\ref{fig:photoz}), neither of which require any dust-obscuration
within \#1916, the latter having the lower formal ${\chi}^2$ value.
We also scale P04's ground-based optical non-detections to that
appropriate for a consistent photometric aperture of three times the
seeing disk at all wavelengths ($V{\ge}26.2$, $R{\ge}26.3$,
$I{\ge}24.7$).  This marginally improves the goodness of fit of the
$z{\simeq}2.6$ solution, and is otherwise indistinguishable from the
original fit.  We retain the matched photometric apertures for the
remainder of this analysis.

To improve the discrimination between low- and high-redshift
solutions, we add the sensitive \emph{HST}/WFPC2 detection threshold
of $R_{702}{\ge}27.0$ (Table~\ref{table:imaging}) to the photometric
constraints.  We fit the synthetic SEDs to the combined
$VRR_{702}IJHK$ dataset.  The goodness of fit of the $z{\simeq}2.59$
solution is marginally worse relative to the $VRIJHK$-band analysis,
because the most stringent optical non-detection comes from the
$R_{702}$-band.  However, in general, the ${\chi}^2$ as a function of
redshift is indistinguishable between this fit and the $VRIJHK$-band
fits.  This is probably because the ${\chi}^2$ is dominated by the
non-detections in six out of the eight observed filters.  To test this
we re-fit the models, limiting the data to just the
$R_{702}JHK$-bands, i.e.\ the most sensitive non-detection
($R_{702}$), the reddest non-detection ($J$) and the two detections
($HK$).  The impact of the sensitive detection limit from the
\emph{HST} data is now clearly evident in the new ${\chi}^2$
distribution, as shown in Fig.~\ref{fig:photoz}.  The only acceptable
fit to the $R_{702}JHK$-band data is at $z{\simeq}10$.

We show the best fit SEDs at $z{=}2.6$ and $z{=}10$, based on the
$R_{702}JHK$-band data in the right panel of Fig.~\ref{fig:photoz}.
This demonstrates the potential power of IRAC photometry to further
discriminate between low and high-redshift interpretations of
candidate high-redshift galaxies.  The SED of low redshift solutions
(e.g.\ $z{\simeq}2.6$) is red at ${\lambda}_{\rm
  obs}{\sim}3{-}5{\mu}{\rm m}$, and the SED of high redshift solutions
($z{\simeq}10$) is blue at similar wavelengths.  We add the detection
limits obtained at ${\lambda}_{\rm obs}{=}3.6{\mu}{\rm m}$ and
$4.5{\mu}{\rm m}$ to the $R_{702}JHK$-band data and re-fit the
spectral templates.  The result is shown in the left panel of
Fig.~\ref{fig:photoz} -- the reduced ${\chi}^2$ of the $z{\simeq}2.6$
solution is now in excess of 2, and the only redshifts which achieve
an acceptable fit to the data are $9.5{\ls}z{\ls}11.5$.

In summary, a single sensitive non-detection derived from \emph{HST}
imaging, combined with deep ground-based near-infrared imaging and
${\sim}2,400$ second integrations with \emph{Spitzer}/IRAC at $3.6$ and
$4.5{\mu}{\rm m}$ can break the degeneracy between low and
high-redshift interpretations of candidate $z{\gs}7$ stellar systems
(see also Egami et al.\ 2004; Eyles et al.\ 2005).  Whilst this is not
an exhaustive study, it demonstrates that despite the demise of
\#1916, searches for gravitationally magnified galaxies at extremely
high redshifts using ground-based near-infrared data remain feasible
when combined with sensitive space-based optical and mid-infrared
data.

\section{Conclusions}\label{conclusions}

We have analyzed new and archival observations of the $z{=}10$ galaxy
\#1916 behind the foreground galaxy cluster lens A\,1835 ($z{=}0.25$)
spanning $0.35{\le}{\lambda}_{\rm obs}{\le}5{\mu}{\rm m}$.
Statistically significant flux is not detected in any of these data,
including our independent re-analysis of P04's discovery $H$- and
$K$-band data.  The $3{\sigma}$ detection thresholds are:
$F({\lambda}_{\rm obs}{=}0.44{\mu}{\rm m}){\gs}4.5{\times}10^{-19}{\rm
  erg\,s^{-1}\,cm^{-2}}$ per spectral resolution element, $R_{702}{\ge}27.0$,
$H{\ge}25.0$, $K{\ge}25.0$, $F(3.6{\mu}{\rm m}){\le}0.75{\mu}{\rm Jy}$ and
$F(4.5{\mu}{\rm m}){\le}0.75{\mu}{\rm Jy}$, where the photometric
limits are calculated in apertures with a diameter $3{\times}$ that of
the seeing disk.

Combining these results with those of Bremer et al.\ (2004) and
Weatherley et al.\ (2004), we are therefore forced by the data to
conclude that there is no statistically sound evidence for the
existence of \#1916.  We also show that inefficient bad-pixel
rejection and issues relating to the calculation of the background
noise can broadly account for the differences between P04's
near-infrared photometry and our own using the same data.  The balance
of probability is therefore that \#1916 was a false detection in P04's
analysis.  From a broader perspective, the demise of \#1916 warns of
the hazards of operating close to the detection threshold of deep
ground-based near-infrared data.

The need for gravitational magnification to boost the observed flux of
faint ($L{\ls}0.1\,L^\star$) galaxies at $z{\sim}6$--8 and populations
of galaxies at still higher redshifts (see \S\ref{intro}) is
undiminished by our results on \#1916.  However an important issue is
whether it is feasible to find such galaxies using ground-based
facilities.  We explore this issue using {\sc hyperz} to fit synthetic
spectral templates to representative data.  Our main conclusions are
that deep near-infrared imaging similar to that presented by P04 in
combination with a single sensitive optical non-detection from
\emph{HST} imaging and moderate depth \emph{Spitzer}/IRAC imaging at
$3.6$ and $4.5{\mu}{\rm m}$ can discriminate between low
(e.g.\ $z{\sim}2{-}3$) and high (e.g.\ $z{\gs}7$) redshift solutions
with strong statistical significance.  In summary, ground-based
near-infrared surveys of massive galaxy cluster lenses with 10-m
class telescopes remain a powerful tool for the discovery of
intrinsically faint galaxies ($L{\ls}0.1L^\star$) at $z{>}7$ that may
be responsible for cosmic re-ionization.  Future surveys should
combine these data with sensitive space-based optical and
mid-infrared observations.

\section*{Acknowledgments}

We acknowledge the bold efforts of Roser Pell\'o and collaborators,
and GPS acknowledges cordial discussions with Roser in Lausanne during
the summer of 2004.  Special thanks go to Jean-Paul Kneib for making
the raw near-infrared imaging data available.  GPS acknowledges the
Caltech Optical Observatories TAC for enthusiastic support, and thanks
Richard Ellis and Avishay Gal-Yam for helpful comments on drafts of
the manuscript.  Thanks also go to Malcolm Bremer, Michael Cooper, Joe
Jensen, Dan Stark, Chuck Steidel and Dave Thompson for a variety of
discussions and assistance.  DJS thanks Dawn Erb, Alice Shapley,
Naveen Reddy and Tommaso Treu for assistance with the LRIS data
reduction, and acknowledges financial support from NASA's Graduate
Student Research Program under NASA grant NAGT-50449.  DS and PRME
acknowledge support from NASA.  We recognize and acknowledge the
cultural role and reverence that the summit of Mauna Kea has within
the Hawaiian community. We are fortunate to conduct observations from
this mountain.


\begin{references}

\reference{} Barger A.J., Cowie L.L., Capak P., Alexander D.M., Bauer
             F.E., Fernandez E., Brandt W.N., Garmire G.P.,
             Hornschemeier A.E., 2003, AJ, 126, 632

\reference{} Becker R.H., Fan X., White R.L., Strauss M.A., Narayanan
             V.K., Lupton R.H., Gunn J.E., Annis J., Bahcall N.A.,
             Brinkmann J., et al., 2001, AJ, 122, 2850

\reference{} Bolzonella M., Miralles J.-M., Pell\'o R., 2000, A\&A,
             363, 476

\reference{} Blain A.W., Chapman S.C., Smail I., Ivison R., 2004, ApJ,
             611, 725

\reference{} Borys C., Chapman S., Donahue M., Fahlman G., Halpern M.,
             Kneib J.-P., Newbury P., Scott D., Smith G.P., 2004,
             MNRAS, 352, 759

\reference{} Bouwens R.J., Thompson R.I., Illingworth G.D., Franx M.,
             van~Dokkum P.G., Fan X., Dickinson M.E., Eisenstein D.J.,
             Rieke M.J., 2004, ApJ, 616, L79

\reference{} Bremer M.N., Jensen J.B., Lehnert M.D.,
             F\"orster-Schreiber N.M., Douglas L., 2004, ApJ, 615, L1

\reference{} Bruzual A.G., Charlot S., 1993, ApJ, 405, 358

\reference{} Bunker A.J., Stanway E.R., Ellis R.S., McMahon R.G.,
             2004, MNRAS, 355, 374

\reference{} Calzetti D., Armus L., Bolin R.C., Kinney A.L., Koornneef
             J., Storchi-Bergmann T., 2000, ApJ, 553, 682

\reference{} Casertano S., de Mello D., Dickinson M., Ferguson H.C.,
             Fruchter A.S., Gonzalez-Lopezlira R.A., Heyer I., Hook
             R.N., Levay Z., Lucas R.A., et al., 2000, AJ, 120 2747

\reference{} Edge A.C., Smith G.P., Sand D.J., Treu T., Ebeling H.,
             Allen S.W., van Dokkum P.G., 2003, ApJ, 599, L69 

\reference{} Egami E., Kneib, J.-P., Rieke G.H., Ellis R.S., Richard
             J., Rigby J., Papovich C., Stark D., Santos M.R., Huang
             J.-S., et al., 2005, ApJ, 618, L5

\reference{} Ellis R.S., Santos M.R., Kneib J.-P., Kuijken K., 2001,
             ApJ, 560, L119

\reference{} Eyles L., Bunker A., Stanway E., Lacy M., Ellis R.S.,
             Doherty M., 2005, MNRAS, submitted, astro-ph/0502385

\reference{} Fan X., Narayanan V.K., Strauss M.A., White R.L., Becker
             R.H., Pentericci L., Rix H.-W., 2002, AJ, 123, 1247

\reference{} Fan X., Narayanan V.K., Lupton R.H., Strauss M.A., Knapp
             G.R., Becker R.H., White R.L., Pentericci L., Leggett
             S.K., Zolt\'an H., et al., 2001, AJ, 122, 2833

\reference{} Fazio G.G., Hora J.L., Allen L.E., Ashby M.L.N., Barmby
             P., Deutsch L.K., Huang J.-S., Kleiner S., Marengo M.,
             Megeath S.T., et al., 2004, ApJS, 154, 10

\reference{} Franx M., Illingworth G.D., Kelson D.D., van~Dokkum P.G.,
             Tran K.-V., 1997, ApJ, 486, L75

\reference{} Hu E.M., Cowie L.L., McMahon R.G., Capak P., Iwamuro F.,
             Kneib J.-P., Maihara T., Motohara K., 2002, ApJ, 568, L75 

\reference{} Ivison R.J., Smail I., Barger A.J., Kneib J.-P., Blain
             A.W., Owen F.N., Kerr T.H., Cowie L.L., 2000, MNRAS, 315,
             209

\reference{} Kneib J.-P., Ellis R.S., Santos M.R., Richard J., 2004,
             ApJ, 607, 697

\reference{} Kneib J.-P., van der Werf P.P., Kraiberg Knudsen K.,
             Smail I., Blain A.W., Frayer D., Barnard V., Ivison R.,
             2004, MNRAS, 349, 1211

\reference{} Kogut A., Spergel D.N., Barnes C., Bennett C.L., Halpern
             M., Hinshaw G., Jarosik N., Limon M., Meyer S.S., Page
             L., Tucker G.S., Wollack E., Wright E.L., 2003, ApJS,
             148, 161

\reference{} Labb\'e I., Franx M., Rudnick G., F\"orster Schreiber
             N.M., Rix H.-W., Moorwood A., van~Dokkum, P.G.,
             van~der~Werf P., R\"ottgering H., van~Starkenburg L., et
             al., 2003, AJ, 125, 1107

\reference{} Lehnert M.D., F\"orster Schreiber N.M., Bremer M.E.,
             2004, ApJ, submitted, astro-ph/0412432

\reference{} Madau P., 1995, ApJ, 441, 18

\reference{} Mellier Y., Fort B., Soucail G., Mathez G., Cailloux M.,
             1991, ApJ, 380, 334

\reference{} Oke J.B., 1990, AJ, 99, 1621

\reference{} Oke J.B., Cohen J.G., Carr M., Cromer J., Dingizian A.,
             Harris F.H., 1995, PASP, 107, 375

\reference{} Pell\'o R., Schaerer D., Richard J., le~Borgne J.-F.,
             Kneib J.-P., 2004, A\&A, 416, L35

\reference{} Richard J., Schaerer D., Pell\'o R., Le~Borgne J.-F.,
             Kneib J.-P., 2003, A\&A, 412, 57 

\reference{} Sand D.J., Treu T., Ellis R.S., 2002, ApJ, 574, L129

\reference{} Sand D.J., Treu T., Smith G.P., Ellis R.S., 2004, ApJ,
             604, 88

\reference{} Sand D.J., Treu T., Ellis R.S., Smith G.P., 2005, ApJ,
             accepted

\reference{} Santos M.R., Ellis R.S., Kneib J.-P., Richard J.,
             Kuijken K., 2004, ApJ, 606, 683

\reference{} Schlegel D.J., Finkbeiner D.P., Davis M., 1998, ApJ, 500,
             525

\reference{} Sharon K., Ofek E.O., Smith G.P., Broadhurst T., Maoz D.,
             Kochanek C., Oguri M., Suto Y., Inada N., Falco E.E.,
             2005, ApJL, 629, 73

\reference{} Smail I., Ivison R.J., Blain A.W., Kneib J.-P., 2002,
             MNRAS, 331, 495

\reference{} Smail I., Smith G.P., Ivison R.J., 2005, ApJ, in press,
             astro-ph/0506176

\reference{} Smith G.P., Kneib J.-P., Ebeling H., Csozke O., Smail
             I., 2001, ApJ, 552, 493

\reference{} Smith G.P., Smail I., Kneib J.-P., Davis C.J., Takamiya
             M., Ebeling H., Czoske O., 2002, MNRAS, 333, L16

\reference{} Smith G.P., Kneib J.-P., Smail I., Mazzotta P., Ebeling
             H., Czoske O., 2005, MNRAS, 359, 417

\reference{} Steidel C.C., Giavalisco M., Pettini M., Dickinson M.E.,
             Adelberger K.L., 1996, ApJ, 462, L17

\reference{} Stern D., Eisenhardt P., Spinrad H., Dawson S.,
             van~Breugel W., Dey A., de~Vries W., Stanford S.A., 2000,
             Nature, 408, 560 

\reference{} Weatherley S.J., Warren S.J., Babbedge T.S.R., 2004,
             A\&A, 428, L29

\reference{} Yan H., Windhorst R., 2004, ApJ, 600, L1

\end{references}
\end{document}